\documentclass[twocolumn,twoside,slac]{revtex4}
\usepackage{graphicx}
\usepackage{fancyhdr}

\begin{document}

\title{A Software Data Transport Framework for Trigger Applications on Clusters}

%

\author{Timm M. Steinbeck, Volker Lindenstruth, Heinz Tilsner, for the ALICE Collaboration}
\affiliation{Kirchhoff Institute of Physics, Ruprecht-Karls-University Heidelberg, Germany}

\begin{abstract}
In the future ALICE heavy ion experiment at CERN's Large Hadron Collider
input data rates of up to 25~GB/s have to be handled by the
High Level Trigger (HLT) system, which has to scale them down to at most 1.25~GB/s
before being written to permanent storage. 
The HLT system that is being designed to cope with these data rates 
consists of a large PC cluster, up to the order of a 1000 nodes,
connected by a fast network. 
For the software that will run on these nodes a flexible data transport
and distribution software framework has been developed. This framework 
consists of a set of separate components, that can be connected via a common
interface, allowing to construct different configurations
for the HLT, that are even changeable at runtime. To ensure 
a fault-tolerant operation of the HLT, the framework 
includes a basic fail-over mechanism that will be further expanded 
in the future, utilizing the runtime reconnection feature of the 
framework's component interface. First performance tests show very 
promising results for the software, indicating that it can achieve
an event rate for the data transport sufficiently high to satisfy
ALICE's requirements.
\end{abstract}

\maketitle

\thispagestyle{fancy}


\section{Background}

ALICE (A Large Ion Collider Experiment) \cite{ALICEWebNorm,ALICEWebPhys,ALICETP} is a heavy ion experiment that 
is being built for the future Large Hadron Collider (LHC) \cite{LHC} at CERN. It is designed
primarily for operation in the collider's heavy ion (HI) mode, but will also
acquire data in proton-proton (pp) mode. Heavy ion mode is characterized by very large
multiplicities of up to 15.000 particles per event, a resulting
maximum event size of about 70~MB, and an allowed data rate into the last trigger stage,
the High Level Trigger (HLT), of up to 25~GB/s, with the largest data contributor being the Time Projection Chamber (TPC). 
Event rates into the HLT are 200~Hz and 1~kHz for the TPC and other participating central detectors
in HI and pp mode respectively. 
In the HLT the complete event data of the participating detectors is available, and 
its task is to perform a full event reconstruction with this data. As for other LHC experiments,
the HLT's architecture is a large farm consisting of the order of a thousand PC nodes running the Linux
operating system and connected via a fast network. Readout data passes through the cluster
in several steps of analysis and merging. 

In the TPC, as an example,
the reconstruction process starts with the raw ADC values that are read out from the detector via the
fiber optical Detector Data Links (DDLs). Each link corresponds to one of 6 sub-sectors,
called patches, of the TPC's 36 sectors, called slices, and is terminated in a PCI card 
through which data from the detector is read out into the HLT system.
During the readout process an FPGA on the PCI card can perform the first stage of 
an analysis together with programs on the node. For instance three dimensional space-points of the charge clusters 
can be determined directly by the FPGA using the ADC values. These event space-points are
then placed into the node's main memory where they are used for further analysis. 
On the same or other HLT nodes tracklets can be calculated from these space-points. For load-balancing purposes this task 
will be distributed among several cluster nodes. 
Once the tracklets have been determined they are sent to nodes in the next stage. Tracklets
that belong to the same sector, and the same event, are sent to the same node in the next stage.
On that node the tracklets are then merged across slice boundaries, to form tracklets for a
whole sector. In the following steps, the tracklets from multiple sectors are sent to the next level 
of nodes, to be merged into larger groups. One sample sequence is to first merge groups
of adjacent sector-sextetts and then merge the tracks from the resulting six sextetts. 

These hierarchical stages of analysis map naturally onto the detector geometry and hierarchy, as the 
preceeding description shows. A second mapping is possible between this hierarchy and the topology used 
for the network connecting the HLT nodes. As a node in one specific stage has its main amount
of communication with the nodes in the directly preceeding and following stages, it does not require full
bandwidth to every node in the cluster. This in turn drastically reduces the requirement 
for the bisection bandwidth of the network in the cluster, and thus also its cost. To retain maximum
flexibility in the network choice the software framework developed for the HLT 
encapsulates network functionality in a separate class library, described below in 
section \ref{Sec:Networking}. 

\section{The Framework}

\subsection{Overview}

For the transport of the event data through the HLT cluster a software framework was needed with two main 
requirements set for it: Flexibility and efficiency. With regard to flexibility, the designed architecture is
a number of independant components that communicate via a common interface. Using this interface
it is possible to plug the components together in different configurations, as required by 
the current conditions, e.g. a test run or pp or heavy ion mode. Even a reconfiguration of the system at runtime, 
with the necessary reconnections of the components, should be supported by the interface.

Concerning efficiency, the primary requirement is a minimal CPU usage during the transfer of data,
both between components on the same and on different nodes. Any CPU cycle used for data transport
is unavailable for analysis, increasing the number of CPUs needed for the HLT, and as a result also its cost.
As a secondary efficiency requirement, the transport of the data should be performed as 
quickly as possible to reduce the latency for an event. This latency reduction is not of prime 
importance as an increased latency can always be compensated by large enough buffer memory and 
memory prices are steadily decreasing with time. 

C++ was chosen as the framework's implementation language, to take advantage of object-oriented programming
capabilities, such as encapsulation and inheritance. Java and other OO languages were not considered, 
as either their basically interpreted nature rendered them unsuitable for the set performance
and efficiency requirements and/or because of their less widespread support and usage.

\subsection{\label{Sec:Interface}The Framework Interface}

For efficiency reasons, the component communication interface as such works only locally on a node. To avoid 
unnecessary copying steps between
the components, shared memory is used for the exchange of data. Descriptors of the data, holding the shared memory
ID and the offset and size of the data block, are transmitted from a data producer to a consumer. The
publisher-subscriber paradigm, also known as producer-consumer-principle, is used for the interface, so that
multiple consumers can attach to one producer. Each subscriber uses a separate set of named pipes for 
communication with the publisher but all access the same data blocks in shared memory. The descriptors
sent from a publisher to its subscribers are thus identical for each subscriber. A descriptor can hold
multiple data blocks of the same event and in addition to the blocks' locations also contains their 
datatypes as well as an indicator of a block data's origin

One implication of this approach of using one data block in shared memory for processing by multiple subscribers, is
that the buffer management has to be done in the publisher. This further implies, that each subscriber has to inform
its publisher when it has finished processing an event so that the publisher can safely reuse the shared memory 
block for another event.

\subsection{\label{Sec:Networking}Network Communication in the Framework}

\begin{figure}
\centering
\resizebox*{0.5\columnwidth}{!}{
\includegraphics{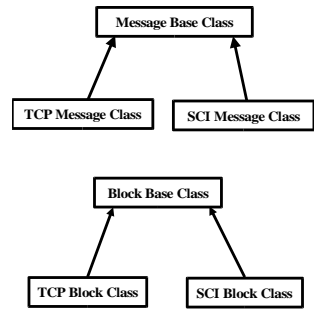}
}
\caption{\label{Fig:ComClasses}Relation of the communication classes}
\end{figure}

As explained above, the interface for communication between the framework components is purely local
on one node. Therefor network communication between components on different nodes has to be handled in 
a different way. Two specialized components have been written, described below in section \ref{Sec:DataFlowComponents},
that work as bridges between nodes. For the network communication these two components make use
of a class library that provides two abstract network communication APIs, one for small
message-like sends and one for large data block transfers. By splitting communication
into these two APIs each of them can be optimized for the corresponding task. Both interfaces
are defined, each in its own abstract base class, so that derived implementation classes
for a specific network technology and protocol can make use of low overhead, low latency, or
high speed features present in that technology. As the call interface is only defined in an abstract
base class it is possible to provide several derived classes with implementations for
each of the two communication types, with each implementation supporting a different network
technology and/or protocol. In the current version of the library, classes for both
communication types are provided for TCP/IP, as the most widely available
baseline network protocol, and SCI network cards by Dolphin \cite{DolphinSCI} using the SISCI API \cite{SISCI},
as an example of a System-Area-Network (SAN). Fig.~\ref{Fig:ComClasses} shows the relation of the different
communication classes in the library. 

\subsection{\label{Sec:FaultTolerance}Framework Fault Tolerance}

One major aspect of any system based on large scale PC clusters, must be the tolerance of the
system as a whole with respect to the failure of hardware or software components on a node,
or even of a whole node. By using the dynamic reconnection feature of the component communication
interface, as well as a number of specialized components, the framework is able to support configurations
that can handle these faults. A number of components are contained in the framework that allow to
setup configurations that handle
faults by distributing workload among a number of nodes. If a component related to one of these nodes fails,
the whole node is deactivated and the load distributed among the remaining nodes. When a stand-by node is available
it can be activated to take over the processing tasks of the deactivated node. A more sophisticated 
system of only temporarily routing a data flow around a fault, restarting and reconnecting failed components
and then again activating the original data flow, is not yet available for the framework, but can be implemented
using external programs.

\subsection{\label{Sec:Components}Framework Components}

Actual functionality of the framework is provided by its components, separate programs that
can be connected together using the interface described in section \ref{Sec:Interface}. A number of 
fully functional components that allow to setup the data flow in a cluster are 
part of the framework. These components, described in section \ref{Sec:DataFlowComponents},
 can be used without change in setups utilizing the framework.
Further components are provided as templates that can be used
to add user specific functionality to a framework system. Three templates are present, for data sources,
data processing components, and data sinks, covered in section \ref{Sec:ComponentTemplates}. 
Adding the appropriate functionality to 
these templates allows to create a framework configuration that is able to address specific 
requirements for a local system. The final group of components presented in section \ref{Sec:FaultToleranceComponents}
addresses the fault tolerance (FT)
abilities described in section \ref{Sec:FaultTolerance}. A number of these components are extended versions of 
the basic data flow components, adding special fault tolerance handling capabilities, while others
are custom components with specific FT functionality.

\subsubsection{\label{Sec:DataFlowComponents}Data Flow Components}

\begin{figure}
\centering
\resizebox*{0.75\columnwidth}{!}{
\includegraphics{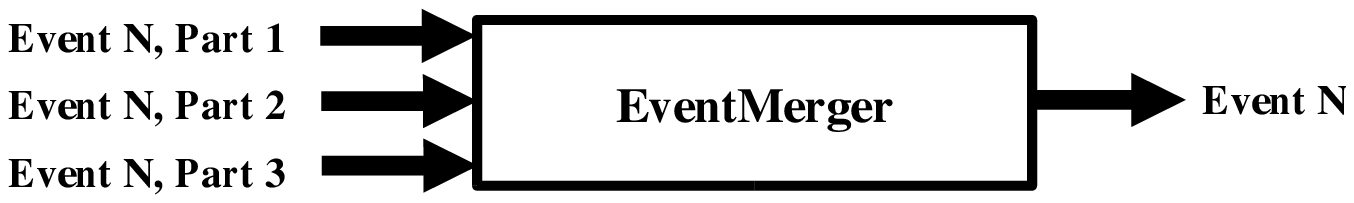}
}
\caption{\label{Fig:EventMerger}The \texttt{EventMerger} component}
\end{figure}

The first data flow component, the \texttt{EventMerger} shown in Fig.~\ref{Fig:EventMerger}, merges multiple data streams,
containing subevents belonging to the same event, into one data stream with larger subevent parts or 
even whole events. For this purpose it waits to receive one subevent from each of its configured input
streams. Once these subevents are received, it constructs a new event descriptor, containing all
data blocks from the received subevents and publishes it for its attached subscribers. 

\begin{figure}
\centering
\resizebox*{0.3\columnwidth}{!}{
\includegraphics{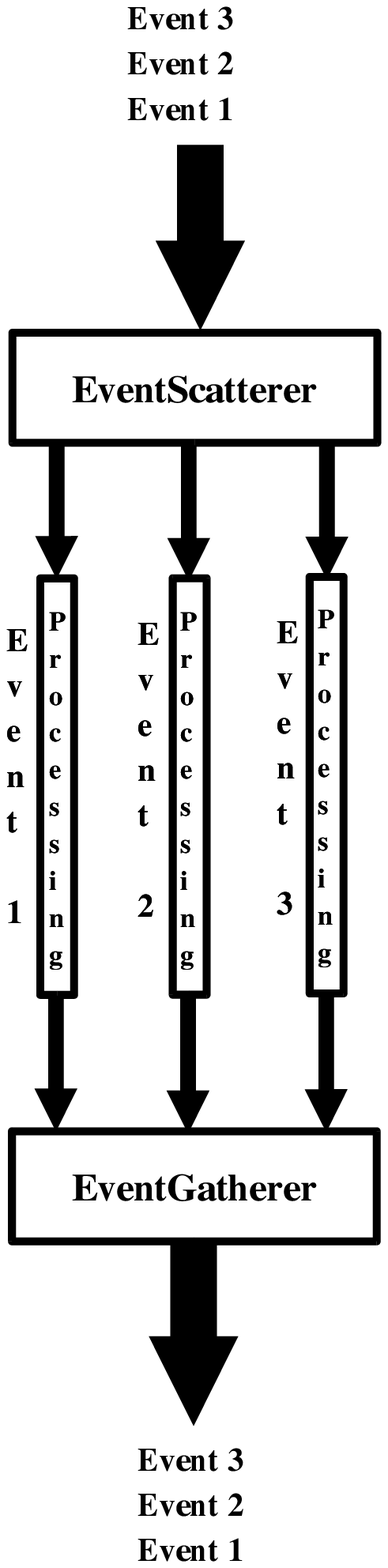}
}
\caption{\label{Fig:EventScatterer-Gatherer}The \texttt{EventScatterer} and \texttt{EventGatherer} components}
\end{figure}

Working in conjunction, the \texttt{EventScatterer} and \texttt{EventMerger} components, shown
in Fig.~\ref{Fig:EventScatterer-Gatherer}, provide a mechanism
to split a single stream of events into multiple smaller event streams, that are later recombined again into
one single stream. This stream splitting can be used for load balancing purposes, with the
resulting multiple streams being distributed among a number of nodes for processing, to be later re-combined.
Splitting up is done on an event-by-event basis, with each single event being dispatched to one output stream 
as a whole. No event is split up into multiple sub-events for distribution. 

\begin{figure}
\centering
\resizebox*{1.0\columnwidth}{!}{
\includegraphics{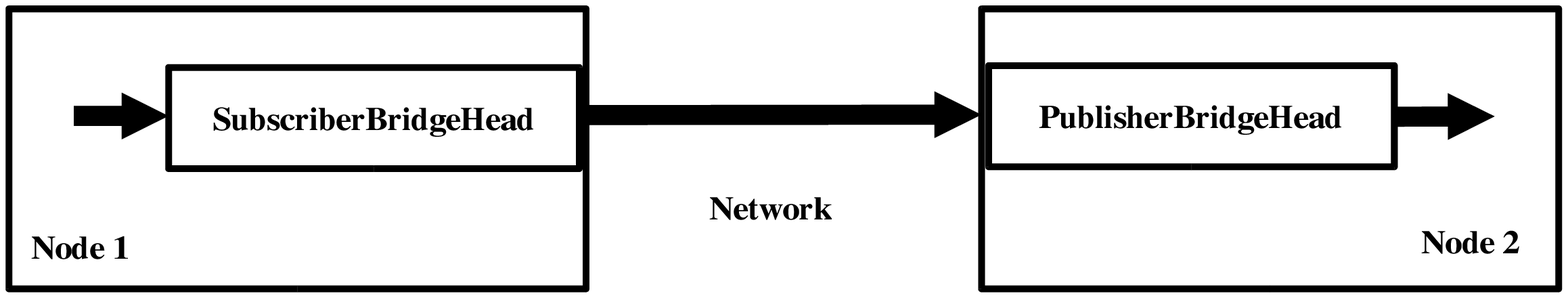}
}
\caption{\label{Fig:BridgeComponents}The \texttt{SubscriberBridgeHead} and \texttt{PublisherBridgeHead} components}
\end{figure}

Another group of components, the \texttt{SubscriberBridgeHead} and \texttt{PublisherBridgeHead} shown in
Fig.~\ref{Fig:BridgeComponents}, provide a mechanism of transparently connecting components on different nodes, as mentioned
in section \ref{Sec:Networking}. The \texttt{SubscriberBridgeHead} component contains a subscriber class to 
accept events using the common communication mechanism for network transmission to the \texttt{PublisherBridgeHead}
on another node. After receiving the event data and the necessary descriptor field subset the \texttt{PublisherBridgeHead}
again makes the data available to other components on its node using an instance of the standard publisher class.
In this way a transparent component connection is enabled across nodes. Components connected to the \texttt{PublisherBridgeHead}
will behave as if they had been subscribed directly to the data's original publishing component. Network communication 
between the two bridge head components is performed using the network class library presented in section \ref{Sec:Networking},
making the bridges independant of a specific networking technology. To use them with a specific networking technology or 
protocol only an appropriate implementation of the library classes is required. No change to the bridge
components themselves is necessary.

\subsubsection{\label{Sec:ComponentTemplates}Component Templates}

User specific tasks related to framework components can basically be grouped into three different
categories: data sources, processing components, and data sinks. Data sources are components that
obtain data from a source outside of the framework, examples of this are special readout devices
like PCI cards with links to detectors. The data source component template already contains the code for the shared memory
buffer management and for the publishing of events. To use the component, only the specific code to access
the data source concerned (so that event data is placed in the shared memory buffers) has to be added. 
If the buffer management code does not meet the requirements of the data source it can also be
easily replaced. 

Processing components are located in the middle of an analysis chain. They accept new data from
data producers, perform analysis steps with the data, mostly producing some new output data, and make
the new output data available to other data consumers. In the analysis template functionalities
for accepting new input data via the framework interface, dereferencing that data in shared memory for access,
shared memory output buffer management, and the publishing of newly created output data are already 
provided. The processing function for the data is called from this gluecode after dereferencing the input data and 
has just to be implemented as needed in order to create a fully functional data processing component. 

Data sinks are components at the end of a processing chain, accepting data that is then forwarded 
to a destination outside of the framework, e.g. simply written to disk
or tape, passed to a DAQ system, or sent to a mass storage system. In the provided sink template the 
code to accept input data from publisher components and dereference and access that data is already provided.
The data output part that has to be implemented is called after the data has been made
accessible inside the component.

\subsubsection{\label{Sec:FaultToleranceComponents}Fault Tolerance Components}

To handle the tolerance with regard to component faults as described in section \ref{Sec:FaultTolerance} 
three new components have been introduced into the framework and four of the data flow components
described in section \ref{Sec:DataFlowComponents} have been enhanced with FT capabilities. Modified versions 
of both bridge components have been enhanced with the abilities to accept commands from external controlling instances.
Primary  commands for the bridges include connect and disconnect commands, as well as the specification of the 
remote bridge components address, which together allow to dynamically establish bridge connections between nodes.

The other two enhanced components are the event scatterer and gatherer components, that have also been 
modified to accept commands from supervising programs. Here the primary commands for both components are
the disabling and enabling of output or input event stream paths. When an ouput path is disabled in the scatterer,
the events that have been sent to that path and not received back as finished, are distributed again among the remaining 
paths. Events that arrive after a path has been disabled are also dispatched only to one of the remaining streams.
In the gatherer component disabling a path causes events that have been received through that path and that 
are released by its subscribers, to not be released directly, but only to be stored as released. As the events for the failed 
path will be sent on another path by the scatterer, they will be received by one of the other input paths in the
gatherer. As soon as this happens, the event release is immediately sent to the publisher at the path from which the 
event was re-received. For events that are still being processed downstream of the gatherer, just the new received
input path is stored instead of the original disabled one.
Together these components ensure that each event that arrives at the scatterer's input passes the gatherer's
output exactly once, as long as there is at least one active event stream connecting them. 

Of the custom built components the first is a simple fault detector component that acts as a monitoring subscriber.
It is attached to a publisher and if no events have been received from that publisher for a specified time it reports
this associated publisher as defective to the second specialized component, the fault tolerance supervisor. This supervisor
component then sends commands to specified targets, informing them of a disabled event path. Targets for this command
are FT scatterer and gatherer components, as well as the bridge manager, the third custom FT component.
In the bridge manager, lists are kept of nodes and their bridge components, associated with the active event streams as well as spare stream nodes. When a stream is 
deactivated the bridge manager looks for an available spare node and if one is found disconnects the bridges connected to the 
deactivated stream. The addresses of the spare nodes are then sent to the bridges, followed by connection commands. As soon as
the bridges report their respective connections to be established, commands are sent to the scatterer and gatherer involved, 
re-activating the path that had failed. After the first events arrive in the fault detection subscriber the status of the path
is changed back to operational in the supervisor component. 

\section{Tests and Benchmarks}

\subsection{ALICE TPC Slice Test}

To test the functionality and performance of the framework an analysis test has been performed on a cluster in
Heidelberg. In this test one slice of simulated proton-proton events of the ALICE TPC,with 25 events piled-up, has 
been processed. Processing has started with the compressed ADC values, similar to the read-out data format, and has
been executed up to merged tracks for the whole slice. The hardware used for the test was a mix of 19 dual SMP Linux PCs,
with CPU speeds of 733~MHz and 800~MHz and 512~MB of RAM per node. For the network
connection between the nodes Fast Ethernet was sufficient due the comparatively small size of the pp events used, between 5~kB and 20~kB 
packed ADC data per patch.
SuSE Linux 7.2 with a 2.4.18 kernel was used as the nodes' operating system. 

\begin{figure*}[t]
\centering
\resizebox*{0.75\textwidth}{!}{
\includegraphics{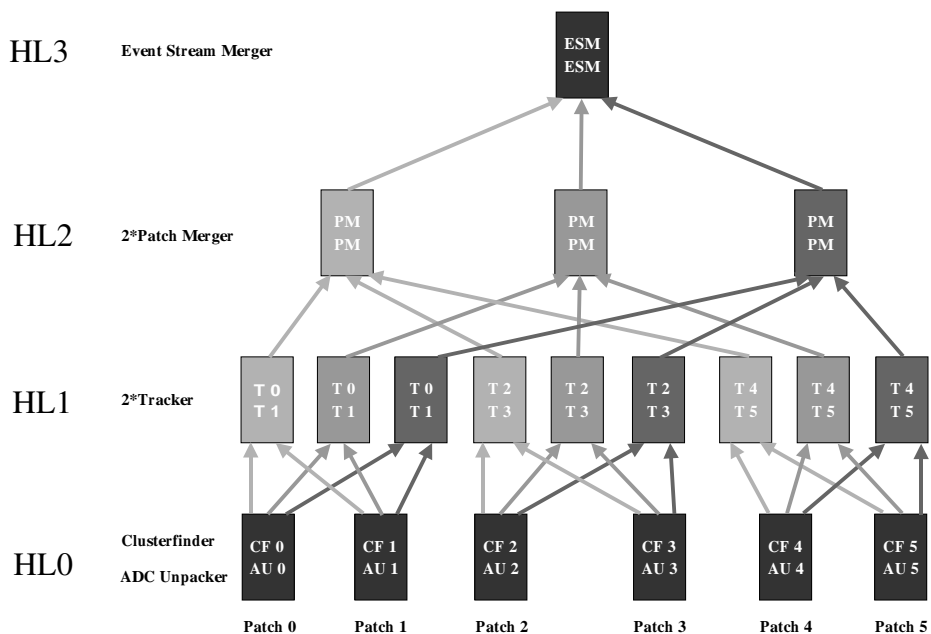}
}
\caption{\label{Fig:SliceTestSetup}The setup used in the 19 node slice analysis test}
\end{figure*}

As shown in the test's software configuration in Fig.~\ref{Fig:SliceTestSetup} the nodes are arranged in four hierarchy levels
(HL), HL0 to HL3. On the six nodes making up the first level the simulated data is read from files and published
as the start of the processing chain. On each node subevents of the same patch in the slice are used for publishing.
The two processing steps performed on the HL0 nodes are the unpacking of the runlength encoded data for processing
and the cluster-finding on the unpacked raw data. Both of these, as well as all following analysis stages, 
are implemented in a separate analysis component so that each processing step could theoretically be executed on a 
separate node. Data of found clusters on one node is then shipped, via an event scatterer and three bridge component pairs, 
to three nodes in the next hierarchy level. 

Of the nine nodes in HL1 three are shared between each pair of adjacent patch nodes in HL0. Cluster data from each
HL0 node is distributed among its three assigned HL1 nodes for track finding. This is done in a way that on each HL1 
node one tracking component runs for each of its two associated patches, to use the nodes two CPUs. The reason
for the even distribution of events among the three nodes, in contrast to e.g. using one node fully for each patch and sharing
only one node between a pair of patches, lies in the better load balancing upon node failures. If one of the HL1 nodes
fails only one third of the processing capability of the concerned patches is lost compared to up to two thirds for
one patch in the other case. 
After tracking, the six patches' tracklets for an event are sent to one of the three nodes in HL2 where track merging on the
slice level is performed. As the last step, the merged slice tracks from the three HL2 nodes are sent to a single event stream
merger node in HL3, collecting all processed events without any additional analysis. The processing components used in the
test are the ones that have already been used for the test in \cite{PARA02Paper}. 

In this test a sustained event processing rate of more than 420~Hz has been achieved. The limiting 
factor was the CPU utilization on the HL0 nodes that was at 100~\% for both CPUs. By splitting up the ADC unpacking and 
cluster finding on separate nodes this bottleneck could be avoided, with the problem in this case, that the unpacked ADC
data constitutes the largest data volume of the different types of data (e.g. packed ADC, unpacked ADC, clusters, 
or tracklets) in the processing. Another approach could be to use a scatterer and a bridge after the file publishing to
send a portion of each patch's packed ADC data for processing to a second HL0 node. 
With the result shown, it has been demonstrated that the framework can be used already today, e.g. in test beams,
and should be ready for both pp and heavy ion mode with their required event rates of 1~kHz and 200~Hz respectively,
given enough processors to perform the required data analysis.

\subsection{Fault Tolerance Test}

\begin{figure}
\centering
\resizebox*{0.75\columnwidth}{!}{
\includegraphics{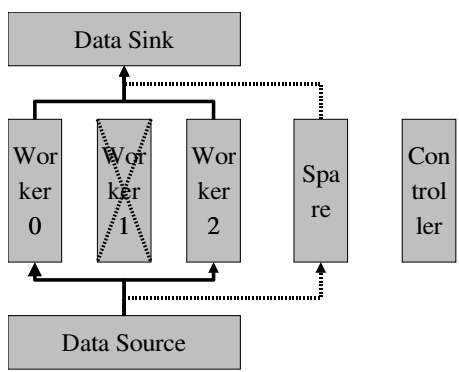}
}
\caption{\label{Fig:FTTest}The fault tolerance test setup}
\end{figure}

As a demonstration of the fault tolerance capabilities contained in the framework a small test was performed with 
a setup of seven nodes similar to the one shown in Fig.~\ref{Fig:FTTest}.
One scattering node sends data, evenly distributed, to three processing nodes that again send their data to a fourth node,
gathering the data. A fifth node is available as a stand-by processing node. The remaining seventh node is used to run the
two control programs, the fault tolerance supervisor and the bridge manager. 

\begin{figure*}[t]
\centering
\resizebox*{0.8\textwidth}{!}{
\includegraphics{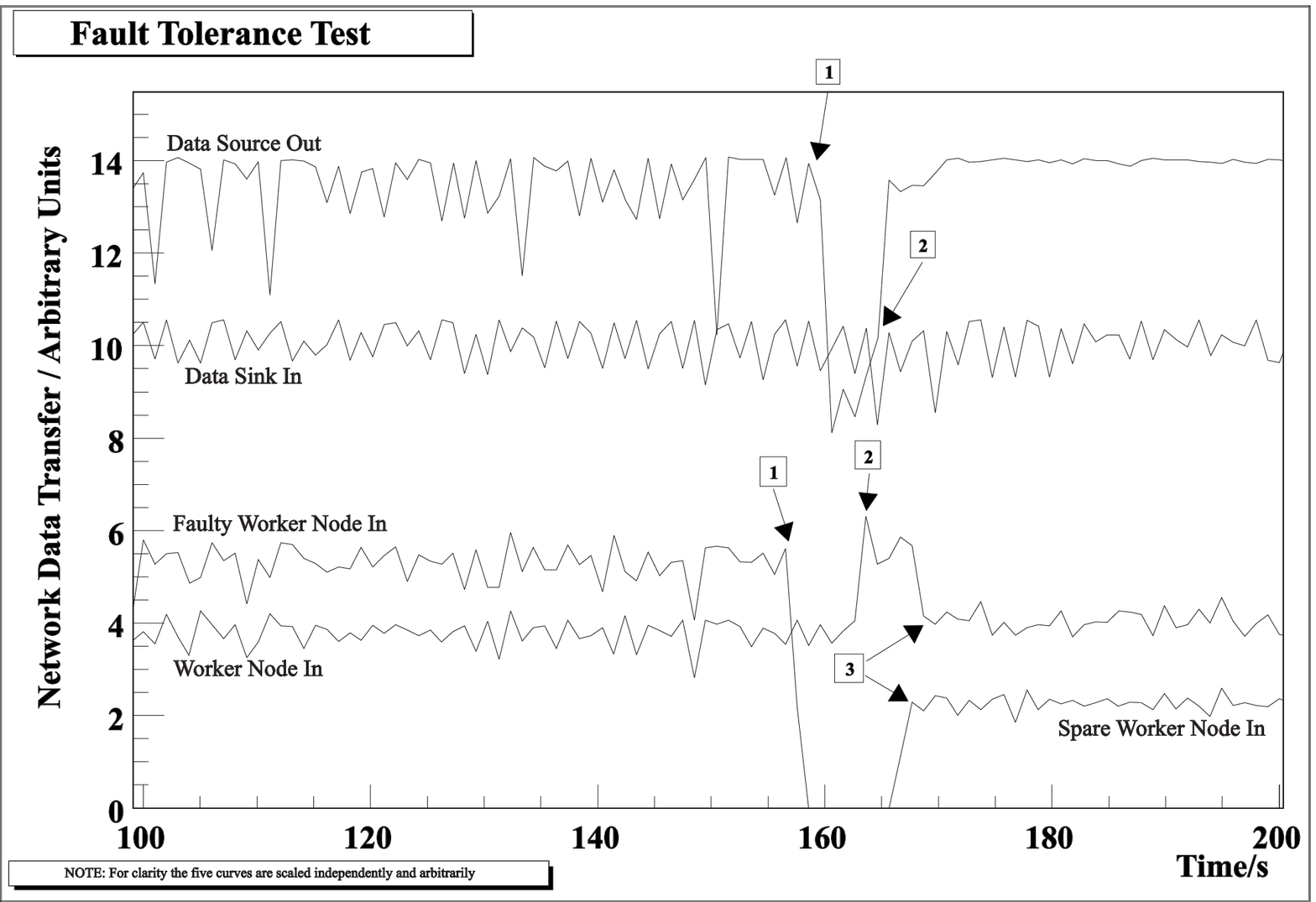}
}
\caption{\label{Fig:FTTestResults}Results from the fault tolerance test. 
The different network throughputs are scaled independantly and arbitrarily.}
\end{figure*}

In the test, the configuration was activated to send 128~kB events at approximately 100~Hz frequency. After a short time 
of running the network cable of one of the three processing nodes was unplugged with the results shown in Fig.~\ref{Fig:FTTestResults}.
The figure shows the network transfers measured during the test on five nodes, the scatterer and gatherer nodes, one working
processing node, the processing node whose network cable was unplugged, and the spare working node. All graphs are
scaled independantly and arbitrarily for visualization purposes, as the absolute measured values are not relevant for this test.

At point 1 in the graph the network cable has been unplugged. The throughput on the worker node involved goes to zero and on
the sending node the throughput drops to about two thirds of its original value, as expected.  Several seconds later, at point 2,
the fault detector has triggered and has informed the tolerance supervisor of the node's ``fault''. The commands to disable
the concerned path between scatterer and gatherer have been sent to those two components. As a result of this action, the 
throughput on the sender increases to its previous value. On the remaining worker shown traffic increases by about 
one-half its previous value, again as expected when the previous workload of three nodes is distributed among two 
remaining ones. Finally at point 3, the bridge manager has replaced the ``faulty'' node with the spare one and reactivated the 
respective path. Network traffic on the spare node is now at the previous value of the removed node and on the remaining
worker the traffic also decreases to its previous value. At that point the operation of the system has been fully restored
to its state prior to the simulated fault. The time elapsed between unplugging and this restoration
is primarily determined by the timeout values chosen for the test, e.g. in the fault detection subscriber. 

With this test it has been shown that the fault tolerance implementation in the framework works and that faults can be handled
by the system already now, although only with a granularity of complete nodes. 

\subsection{Framework Interface Performance}

As a benchmark for the performance and scaling behaviour of the interface used for communication between the framework's
components, a test has been run on three separate dual SMP PCs, with CPU speeds of 733~MHz, 800~MHz, and 933~MHz. 
In the test a publisher announces a continuous stream of $50 \times 10^6$ empty events to a subscriber and the subscriber immediately
releases each received event. Since the events are empty, no event data is used or copied in the test, restricting the
test to measure the interface's performance. 

\begin{table}[hbt]
\caption{\label{Tab:PubSubAverageRates}Average event rates and resulting time overheads.}
\begin{tabular}{|l||c|c|c|}
\hline
& 733 MHz & 800 MHz & 933 MHz  \\
\hline \hline
Average event & 11.86 & 12.73 & 14.41 \\
rate / kHz &  &  &  \\
\hline
Average time & 168.7 & 157.1 & 138.8 \\
overhead / $\mu$s &  &  &  \\
\hline
\end{tabular}
\end{table}

\begin{figure*}[t]
\centering
\resizebox*{0.8\textwidth}{!}{
\includegraphics{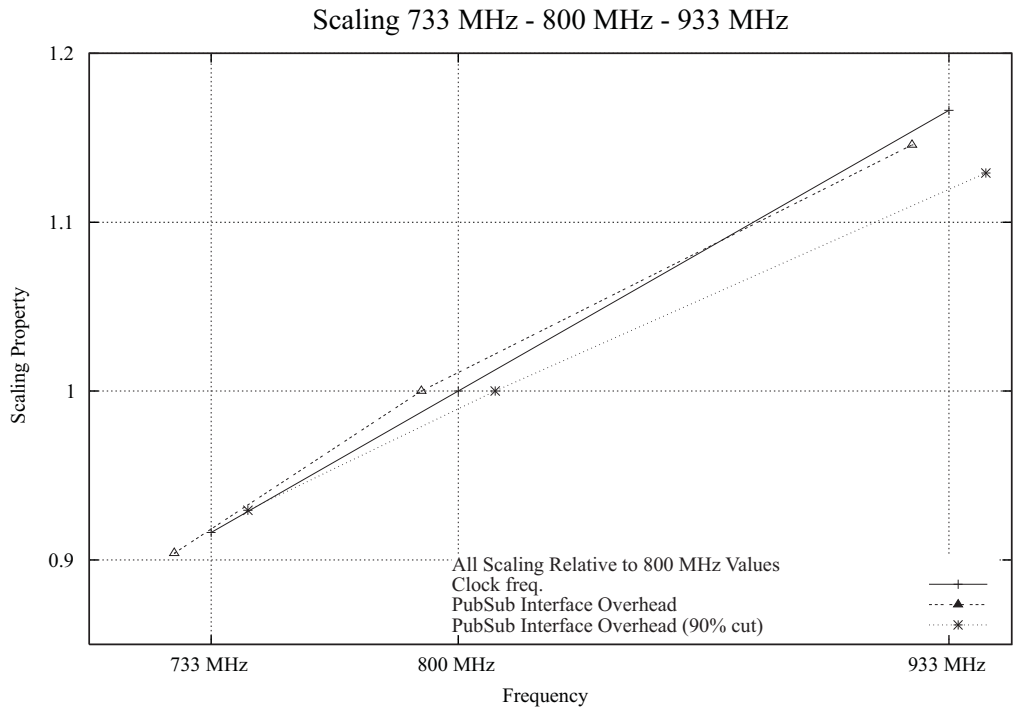}
}
\caption{\label{Fig:ScalingResults}The scaling behaviour of the framework interface}
\end{figure*}

The results that have been obtained from this test are given in Table~\ref{Tab:PubSubAverageRates} and
Fig.~\ref{Fig:ScalingResults}. In the table, the achieved average event rates over the test's running time are shown 
together with the corresponding time overhead. Fig.~\ref{Fig:ScalingResults} shows two different scaling behaviours, 
with and without a cut, obtained during the benchmark together with the clock speed scaling as a point of reference. 
Application of the cut was performed in order to use the timing obtained from the fastest 90~\% of events only. The
motivation for this is that programs get descheduled during the test by the operating system. Since the timing is measured
by reading the current system time at the beginning and end of multiple involved function blocks, a descheduling
in such a block will increase the measured time incorrectly. As these descheduling events do not cause
a fixed time increment the cut was made to exclude the 10~\% of events that took the longest time to process
in each of the function blocks and thereby remove the influences of descheduling events on the interface performance
measurements.
In the table
the cut is not included so that these numbers indicate the absolute rate that can be achieved in an actual system. 
As can be seen from the table and the figure, the interface scales reasonably well with the clock frequency.
It does not scale perfectly but the framework should be able to take advantage of a good fraction
of future increases in CPU clock frequency making it appropriate for use on CPUs expected for the time
when the ALICE HLT starts to operate. Recent tests with a number of performance enhancements in the interface code
have indicated that, for future versions, a performance increase of about a factor of four can be expected.

\section{Conclusion and Outlook}

We have presented a working framework for constructing distributed online data analysis chains running on
Linux clusters. The framework allows a flexible configuration due to its approach of components
that can be connected together via a defined common interface. Using a separate class library
to encapsulate network communication makes the framework and its components independant of 
the actual network technology and protocol used, allowing for more flexibility in the construction
of a cluster. With the described tests the framework has been shown to work and to be usable
in real applications already today, provided that enough CPU power is available to perform the 
desired analysis which is independant of the framework as such. 
Future work on the framework has to include a simple method to specify the configuration to
be used in a cluster as well as an efficient method of starting and supervising the involved 
processes on the cluster nodes. This supervision, in principle a detector control system, 
also involves a central or distributed fault tolerance control and decision unit that 
should allow for a finer grained control of the fault tolerance and recovery actions. Finally,
more tuning measures on the framework, at least completing the above mentioned preliminary performance 
enhancements, will be done as well.

\begin{acknowledgments}

Work on the ALICE High Level Trigger has been financed by the German Federal Ministry of Education and Research (BMBF)
as part of its program ``F\"orderschwerpunkt Hadronen- und Kernphysik - Gro\ss{}ger\"ate der physikalischen Grundlagenforschung''.

\end{acknowledgments}


\begin{thebibliography}{9}   




\bibitem{ALICEWebNorm} 
ALICE homepage: 

\texttt{http://alice.web.cern.ch/ALICE/},
May 2003

\bibitem{ALICEWebPhys} 
ALICE pages for physicist: 

\texttt{http://alice.web.cern.ch/ALICE/user.html}, 
May 2003

\bibitem{ALICETP}
The Alice Collaboration, ``ALICE - Technical Proposal for A Large Ion Collider Experiment at the CERN LHC'',
CERN/LHCC/95-71, LHCC/P3, Dec. 15th 1995

\bibitem{LHC}
LHC Project Homepage: 

\texttt{http://lhc-new-homepage.web.cern.ch/}\textbackslash

\texttt{lhc-new-homepage/}, 
May 2003

\bibitem{DolphinSCI}
Dolphin Homepage:

\texttt{http://www.dolphinics.com/}, 
May 2003

\bibitem{SISCI}
F. Giacomini et al., ``Low-level SCI software functional specification'',
Esprit Project 23174 - Software Infrastructure for SCI (SISCI)

\bibitem{PARA02Paper}
T. M. Steinbeck et al., ``A Framework for Building Distributed Data Flow Chains in Clusters'',
in Lecture Notes in Computer Science LNCS 2367, 
Proceedings of the 6th International Conference on Applied Parallel Computing, PARA 2002, 
Springer Verlag Berlin Heidelberg, 2002

\end{thebibliography}

\end{document}